\documentclass[%
 11pt,
 amsmath,amssymb,twocolumn
]{article}

\usepackage{wrapfig}
\usepackage{authblk}
\usepackage{verbatim}

\usepackage[
    style=numeric, % in-between (includes titles)
    citestyle=numeric,
    sorting=none, % sort the bib by order referenced
]{biblatex}
\usepackage[utf8]{inputenc}
\usepackage[english]{babel}
\usepackage{url}
\usepackage[margin=0.95in]{geometry}
\usepackage{graphicx}% Include figure files

\usepackage{hyperref}
\usepackage[dvipsnames]{xcolor} %get other color names
\hypersetup{
    colorlinks=true,
    citecolor=black,     
    urlcolor=RawSienna,
    linkcolor=BlueViolet,
}

\usepackage{dcolumn}% Align table columns on decimal point
\usepackage{bm}% bold math
\usepackage{titlesec}
\usepackage[font=small,labelfont=bf]{caption}
\usepackage[belowskip=-15pt,aboveskip=0pt]{caption}

\usepackage{enumitem}

\usepackage{array}
\newcolumntype{L}[1]{>{\raggedright\let\newline\\\arraybackslash\hspace{0pt}}m{#1}}
\newcolumntype{C}[1]{>{\centering\let\newline\\\arraybackslash\hspace{0pt}}m{#1}}
\newcolumntype{R}[1]{>{\raggedleft\let\newline\\\arraybackslash\hspace{0pt}}m{#1}}

\usepackage[font={small,it}]{caption}

\titlespacing*{\section}
{0pt}{1ex}{1ex}
\titlespacing*{\subsection}
{0pt}{1ex}{1ex}
\titleformat*{\section}{\large\bfseries}
\titleformat*{\subsection}{\normalsize\bfseries}

\begin{document}

\renewcommand{\thefootnote}{\alph{footnote}}	

\begin{titlepage}
    \begin{center}
    \vspace*{3.0cm}
    \LARGE
    \textbf{Advancing the Scientific Frontier with Increasingly Autonomous Systems} 
    \\ 
    \vspace{4.0cm}
    \normalsize
    Rashied Amini$^{1,}$\footnote{
    \href{mailto:rashied.amini@jpl.nasa.gov}{rashied.amini@jpl.nasa.gov}, (626) 720-9942
    \par\textit{\copyright 2020 California Institute of Technology. This work was carried out at the Jet Propulsion Laboratory, California Institute of Technology, under a contract with the National Aeronautics and Space Administration (80NM0018D0004).}},
    Abigail Azari$^{2}$,
    Shyam Bhaskaran$^{1}$,
    Patricia Beauchamp$^{1}$,
    Julie Castillo-Rogez$^{1}$,
    Rebecca Castano$^{1}$,
    Seung Chung$^{1}$,
    John Day$^{1}$,
    Richard Doyle$^{1}$,
    Martin Feather$^{1}$,
    Lorraine Fesq$^{1}$, 
    Jeremy Frank$^{3}$,
    P. Michael Furlong$^{3}$, 
    Michel Ingham$^{1}$, 
    Brian Kennedy$^{1}$,
    Ksenia Kolcio$^{4}$,
    Issa Nesnas$^{1}$,
    Robert Rasmussen$^{1}$,
    Glenn Reeves$^{1}$,
    Cristina Sorice$^{1}$,
    Bethany Theiling$^{5}$,
    Jay Wyatt$^{1}$\\
    
\begin{comment}    
    \begin{figure}[h!]
        \centering
        \includegraphics[width=0.7\textwidth]{CoverArt.png}
    \end{figure}
\end{comment}

    \vspace{1.0cm}
    \large
    \textbf{Endorsements}\\
    \normalsize
    Tibor Balint$^{1}$,
    Luther Beegle$^{1}$,
    Steve Chien$^{1}$,
    Chad Edwards$^{1}$,
    Carolyn Ernst$^{6}$,
    Terrance Fong$^{3}$,
    Abigail Fraeman$^{1}$,
    Anthony Freeman$^{1}$,
    Carly Howett$^{7}$,
    Robert Lillis$^{2}$,
    Gentry Lee$^{1}$,
    Michael Mischna$^{1}$,
    Marc Rayman$^{1}$,
    Mark Robinson$^{8}$
    \\
    \vspace{1.0cm}
    \textbf{September 15, 2020}\\
    \vspace{1.0cm}
    \normalsize
    $^{1}$\textit{NASA Jet Propulsion Laboratory}\\
    $^{2}$\textit{Space Sciences Laboratory, University of California, Berkeley}\\
    $^{3}$\textit{NASA Ames Research Center}\\
    $^{4}$\textit{Okean Solutions}\\
    $^{5}$\textit{NASA Goddard Space Flight Center}\\
    $^{6}$\textit{Johns Hopkins University Applied Physics Laboratory}\\
    $^{7}$\textit{Southwest Research Institute}\\
    $^{8}$\textit{Arizona State University}\\
    \end{center}
\end{titlepage}

%\maketitle
%\thispagestyle{empty}
\clearpage

\section{\label{sec:intro}Executive Summary}
A close partnership between people and partially autonomous machines has enabled decades of space exploration. But to further expand our horizons, our systems must become more capable. Increasing the nature and degree of autonomy - allowing our systems to make and act on their own decisions as directed by mission teams - enables new science capabilities and enhances science return. The 2011 Planetary Science Decadal Survey (PSDS) and on-going pre-Decadal mission studies have identified increased autonomy as a core technology required for future missions. However, even as scientific discovery has necessitated the development of autonomous systems and past flight demonstrations have been successful, institutional barriers have limited its maturation and infusion on existing planetary missions. Consequently, the authors and endorsers of this paper recommend that new programmatic pathways be developed to infuse autonomy, infrastructure for support autonomous systems be invested in, new practices be adopted, and the cost-saving value of autonomy for operations be studied. 

\section{\label{sec:motivations}Motivations}
From the beginning of interplanetary exploration, reliance on on-board decision-making has been critical for mission success. The use of time-driven command sequences and critical sequence retries on the Viking and Mariner 6 \& 7 orbiters resulted in reduced risk and increased the ability of spacecraft to return science data \parencite{mariner1971, brown_1997}. Over the following decades, the evolutionary inclusion of autonomous functions, primarily in the domains of spacecraft fault protection, and guidance and control, further reduced mission risk and increased a spacecraft's ability to perform and downlink science measurements. For instance, the Jovian radiation environment caused multiple safe mode events during the Galileo mission; but ``smart safing'' enabled it to maintain thermally-safe attitude to protect its instruments despite immediate loss of operator control. However, the mission and system complexity needed to answer new questions in planetary science has outpaced efforts to mature autonomous capabilities. As detailed in this paper, many of the ambitious mission concepts described in the Planetary Mission Concept Studies (PMCS) for the 
Planetary Science and Astrobiology Decadal Survey (PSADS) and the 2018 Workshop on Autonomy for Future NASA Science Missions will not be achievable using the current paradigm of spacecraft control \cite{Planetar61,AutonomyWorkshop2018}. 

To answer ever more challenging science questions, we will need spacecraft that can explore unknown and dynamic environments with less input from human operators. This will demand an integrated approach to autonomy, because autonomy is a system-level technology that requires an interdisciplinary approach to technology development. The need for a revolutionary advance in spacecraft autonomy to meet the needs of NASA’s missions was identified as early as the 1980 Carl Sagan, et al. report to NASA on Machine Intelligence and Robotics \cite{sagan1980machine}. Yet despite the need for and past uses of autonomous capabilities, these capabilities are rarely used or even considered in most planetary missions due to institutional barriers.  

In the Sagan Report, one of the primary barriers to progress in autonomous systems was identified as culture: \textit{``Technology decisions are, to much too great a degree, dictated by specific mission goals, powerfully impeding NASA utilization of modern computer science and technology. Unlike its pioneering work in other areas of science and technology, NASA’s use of computer science and machine intelligence has been conservative and unimaginative.''} Similar cultural issues were identified at ESA, based on the experience of 2019 OP-SAT mission: \textit{``...resistance to experimentation and innovation, especially when the projected benefits are not yet flight proven. Time after time projects settle for reuse rather than innovation.''} \cite{evans2016ops}  To wit, the highest level recommendation of this paper is one made by the Sagan Report: \textit{``The advances and developments in machine intelligence and robotics needed to make future space missions economical and feasible will not happen without a major long-term commitment and centralized, coordinated support.''}

\section{The Need for Autonomy}
\begin{figure}
    \centering
    \includegraphics[width=0.45\textwidth]{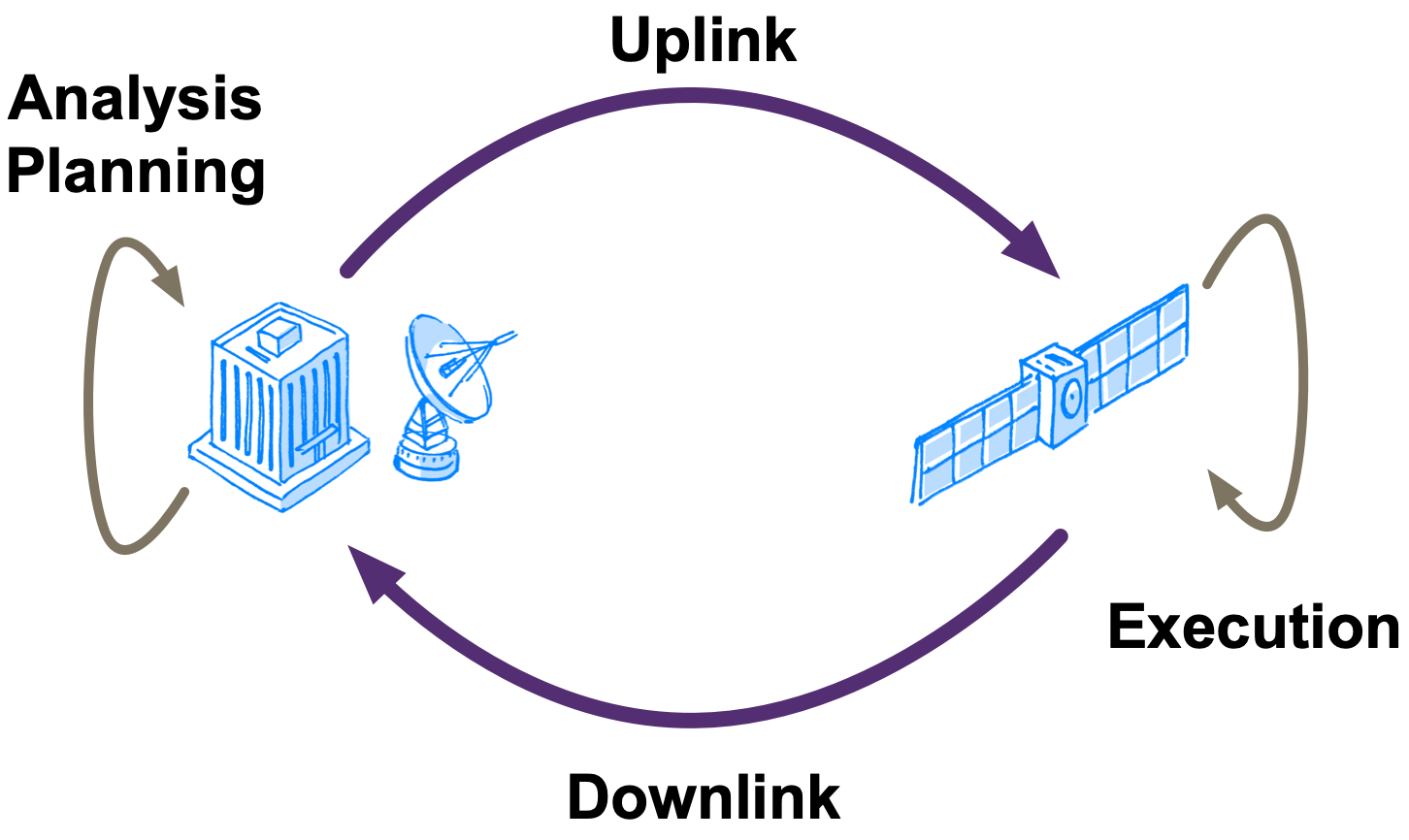}
    \caption{The traditional commanding control loop has a characteristic timescale that may be greater than the timescales needed to exercise control. In these cases, autonomous functions are required.}
    \label{fig:ControlLoop}
\end{figure}

The 2015 NASA Technology Roadmap defines autonomy as ``the ability of a system to achieve goals while operating independently of external control.'' \cite{taxonomy} External control is exemplified by the traditional commanding control loop, summarized in Figure \ref{fig:ControlLoop}. Autonomy reduces the characteristic control timescale by moving analysis and planning on board the spacecraft \cite{chien2017robotic}. Moving functions traditionally performed by mission teams to on board the spacecraft requires changes to both flight systems and ground systems.

When the timescale associated with traditional commanding becomes larger than timescales required for responding to spacecraft critical events or interacting in an unpredictable environment, using {\em a priori} planning for control increases mission risk. Deploying autonomy, {\em in situ}  perception, reasoning, and acting under both nominal and off-nominal situations, mitigates this risk and allows the spacecraft to make decisions based on current circumstances.

Commanding timescales are partly driven by communications for commands and status. For instance, in the case of planetary rotorcraft like Ingenuity or Dragonfly, communication is constrained by limited data rates (e.g. due to weak signal), restricted link availability (e.g. a body’s orbit and day/night cycle), and large data products required for ground-based planning (e.g. contextual maps for path planning) \cite{lorenz2018dragonfly}. The timescale of systems responding to critical events or performing critical operations is driven by the dynamical nature of system-environment and system-system interactions and how predictable and observable these interactions are. Dynamic rotorcraft mobility in planetary atmospheres is incompatible with communication constraints and has to occur {\em in situ}. 

\begin{figure}
    \centering
    \includegraphics[width=0.45\textwidth]{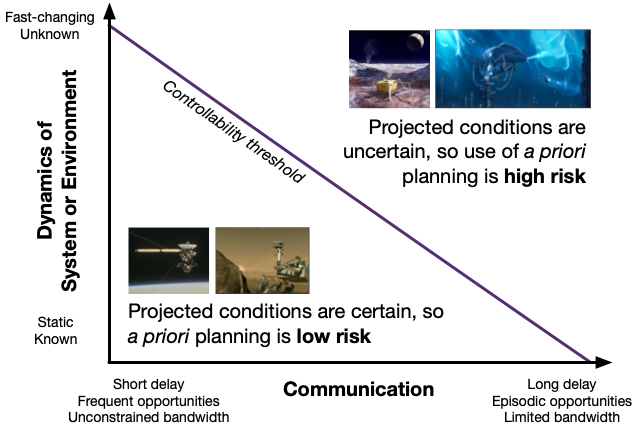}
    \caption{Autonomy is necessary when a spacecraft has to react to changes in the environment or within itself at a shorter timescale than afforded by communication constraints. Past missions relied on their ability to predict but future missions that operate in increasingly unknown environments would fall outside the controllability threshold.}
    \label{fig:AutonomyNeed}
\end{figure}

It is this relationship between communications- and dynamical-related timescales that has traditionally driven requirements for autonomy. Where the communications timescale exceeds the dynamical timescale, safe control using strict {\em a priori} planning cannot be assured and autonomous functions are required for safe control. Figure \ref{fig:AutonomyNeed} illustrates the control regimes and associated risks of using {\em a priori} planning.  It’s for this reason that autonomous functions were first used for mission aspects with short dynamical timescales, e.g. guidance, control, and fault protection \cite{brown_1997}. With incremental confidence from these early missions, more ambitious autonomous functions were flown to enable more ambitious science measurements. For example, in 2004 Deep Impact’s Impactor spacecraft required autonomous navigation functions (AutoNav) to intercept Comet Tempel 1 \cite{kubitschek2006deep}. This was driven by two primary goals: first, colliding with the comet using the Impactor spacecraft, and second, observing the impact on the Flyby vehicle. Because the exact size, shape, and orbit of the comet could not be determined in advance from the ground, closing the navigational loop on-board was necessary to successfully impact the 7 km size comet at a speed of 10 km/s. The use of AutoNav was unavoidable to meet the science objective and AutoNav’s decision-making could be sufficiently modeled and supervised that it could be trusted. 

While using autonomy may be enabling for some missions, it can enhance mission productivity and science return for nearly all missions, increasing a mission's science-to-dollar ratio. On-board science data analysis has been used on various Mars rovers \cite{castano2008automatic,estlin2012aegis} and on Earth Observing-1 \cite{casper} to improve science return and dynamically target and image novel signatures. For instance, Curiosity incorporated Autonomous Exploration for Gathering Increased Science (AEGIS) to target the ChemCam laser-induced breakdown spectroscopy (LIBS) instrument, targeted preferred outcrop terrain over 93\% of the time as opposed to blind targeting, which targeted outcrops 24\% of the time \cite{francis2017aegis}. Two additional white papers submitted to the PSADS, Azari et al and Theiling et al, further describe the advantages of using science autonomy \cite{azari2020integrating, BethanyAutonomy}. 

However, a lesson learned from Mars rovers, EO-1, Deep Impact, and Stardust is that flight-proven and high value-to-risk-ratio autonomous functions are not regularly used on Flagship or competed missions. Moreover, these deployments have not led to the development of system-level autonomy, which is able to integrate numerous autonomous and traditional functions. \textbf{Given NASA’s track record of developing and deploying autonomous systems and the interaction of autonomous functions necessary for strategic planetary science missions, it is clear a shift in institutional culture is necessary.}

\subsection{Future Planetary Missions and Their Autonomy Requirements}
All of the missions under evaluation by the PSADS are either enabled by, or would be enhanced by, the use of increasingly autonomous systems. By evaluating aspects of mission architecture that constrain communications or are driven by the dynamics of system-system and system-environment interactions we can identify how these missions are enabled and enhanced. These evaluations are summarized in Table \ref{tab:tab}, and additional description for two mission types are included below.

\begin{table*}[t]
    \caption{Summary of how different aspects related to mission architecture can be enabled or enhanced through the use of integrated autonomous functions.}
    \label{tab:tab}
\resizebox{\textwidth}{!}{%
\begin{tabular}{|l|l|l|}
\hline
\multicolumn{1}{|c|}{\textbf{Aspect of Mission Architecture}}                                                                                                                                                                                                                                           & \multicolumn{1}{c|}{\textbf{Drivers}}                                                                                                                                                             & \multicolumn{1}{c|}{\textbf{\begin{tabular}[c]{@{}c@{}}Enabling/Enhancing \\ Autonomous Functions\end{tabular}}}                                                                                                             \\ \hline
\begin{tabular}[c]{@{}l@{}}\textbf{Entry, Descent, and Landing}, e.g.\\ - Mercury Lander, Enceladus Orbilander\\\textbf{Surface \& Aerial Mobility}, e.g.\\ - Intrepid Rover\\ - Lunar/Vesta Geochronology Hopper\\ - Dragonfly, Mars Helicopter\end{tabular}                                                                                                  & \begin{tabular}[c]{@{}l@{}}- Short dynamic timescales\\ - Limited {\em a priori} atmosphere/surface\\ characterization\\ - One-way light times (OWLTs)\end{tabular}                                       & \begin{tabular}[c]{@{}l@{}}- Terrain-relative navigation (TRN) \\and dynamic control\\ -  Hazard assessment and avoidance \\ - Payload data analysis\\ - Planning/execution\\ - Restorative fault management\end{tabular} \\ \hline
\begin{tabular}[c]{@{}l@{}}\textbf{Short-Lived Landers}, e.g.\\ - Venus Flagship lander\\ - Europa Lander\\ - VIPER Rover\end{tabular}                                                                                                                                                                           & \begin{tabular}[c]{@{}l@{}}- Science competitiveness\\ - Limited lifetime\\ - Limited bandwidth and contact opportunity\end{tabular}                                                              & \begin{tabular}[c]{@{}l@{}}- Payload data analysis\\ - Planning/execution\\ - Restorative fault management\end{tabular}                                                           \\ \hline
\begin{tabular}[c]{@{}l@{}}\textbf{Missions with Opportunistic Science }\\Potentially all missions, but particularly:\\ - Fast flyby missions, e.g. \cite{DonitzOort}\\ - Monitoring missions, e.g. MOSAIC\\ - In-situ missions, e.g. Ceres Habitability\end{tabular}                                        & \begin{tabular}[c]{@{}l@{}}- Limited communications and time to \\perform critical targeting\\ - Science competitiveness and impact\\ - Cost/risk reduction\end{tabular} & \begin{tabular}[c]{@{}l@{}}- Autonomous navigation (enabling \\for fast flybys)\\ - Planning/execution\\ - Science planning\\ - Payload data analysis\end{tabular}                  \\ \hline
\begin{tabular}[c]{@{}l@{}}\textbf{Interplanetary Cruise}, e.g.\\ - All missions beyond Earth-Moon system\\ - Missions using electric propulsion, e.g. Persephone\end{tabular}                                                                                                      & \begin{tabular}[c]{@{}l@{}}- Mitigating impact of safing on trajectory \\(EP missions)\\ - Cost/risk reduction\end{tabular}                                                                         & \begin{tabular}[c]{@{}l@{}}- Autonomous navigation\\ - Planning/execution\\ - Science planning\\ - Payload data analysis\end{tabular}                  \\ \hline
\begin{tabular}[c]{@{}l@{}}\textbf{Missions with Coordinated Observations}, e.g.\\ - Multi-SC mapping missions, like MOSAIC\\ - Bistatic radar experiments, e.g. CONCERT\\ - Planetary defense \& impactor/observer missions, e.g. \\DART, Small Bodies DRM \cite{SmallBodiesDRM}\\ - Landed system coordination, e.g. M2020/Mars \\Helicopter, lava tube exploration\end{tabular} & \begin{tabular}[c]{@{}l@{}}- Limited time to coordinate with ground-in-\\the-loop\\ - Science competitiveness\\ - Cost/risk reduction\end{tabular}                                                  & \begin{tabular}[c]{@{}l@{}}- Multi-agent coordination\\ - Planning/execution\\ - Science planning\\ - Autonomous navigation\end{tabular}                                          \\ \hline
\begin{tabular}[c]{@{}l@{}}\textbf{Mapping Missions}, e.g.\\ - Europa Clipper\\ - MORIE\end{tabular}                                                                                                                                                                                                             & \begin{tabular}[c]{@{}l@{}}- Mitigating the impact from anomaly\\ - Science competitiveness\\ - Cost/risk reduction\end{tabular}                                                                  & \begin{tabular}[c]{@{}l@{}}- Planning/execution\\ - Science planning\end{tabular}                                                                                                 \\ \hline
\begin{tabular}[c]{@{}l@{}}\textbf{Operations in High-Radiation Environments}, e.g.\\ - Europa Clipper\\ - Io Volatiles Explorer\end{tabular}                                                                                                                                            & \begin{tabular}[c]{@{}l@{}}- Communications delay to restore science \\operations may not be acceptable\\ - Reduce risk/cost\end{tabular}                                                           & - Restorative fault management                                                                                                                                                    \\ \hline

\begin{tabular}[c]{@{}l@{}}\textbf{Approach/Rendezvous with Unexplored Bodies}, e.g.\\ - NEO/NEA missions \cite{SmallBodiesDRM} \\ - Comet Sample Return \\ - Transneptunian bodies/KBO etc.\end{tabular}                                                                                                                                         & \begin{tabular}[c]{@{}l@{}} - Uncertainty in relative spacecraft/body position \\ - Unknown irregular body shape and gravity \\ - Unknown geotechnical properties for landing \\ - Limited {\em a priori} surface characterization\end{tabular}                                                           & \begin{tabular}[c]{@{}l@{}}- Autonomous navigation \\ - Autonomous mapping \\ - TRN, hazard assessment, landing \\ - Autonomous surface navigation \\ - Restorative fault management\\  \end{tabular}                                                                                                                    \\ \hline
\end{tabular}
}
\end{table*}

\textit{Landed Missions (landers, hoppers, rovers, aerial systems)} Landed systems are impacted by similar issues that drive the use of autonomy. Across these classes of missions, the dynamic aspects of operations during entry, descent, and landing (EDL), roving, flying, and hopping require situational awareness and {\em in situ} reasoning and acting. Limited line of sight to Earth and communication constraints restrict contact opportunities and data rates. Limited lifetime and data volume constraints, such as a Europa or a Venus lander, will need situational awareness, assessment of the interaction with the unknown surface for sampling, and ``smart'' targeting to increase science return and reduce risk under off-nominal conditions \cite{wagstaff2019enabling}. On-board analysis of ``remote'', low-cost data, like Raman or LIBS, can significantly reduce risk of performing high-cost sampling, like drilling \cite{ReevesEuropa,LanderUncert,VenusFlagship}.

Just as critical to these time-limited missions is a capability for restorative fault management that can restore functionality after a safing event, or take action to avoid the need for a safing event. As more decision-making capability is moved on-board, the scope of autonomous fault detection, isolation and recovery (FDIR) functions will increase, relying on on-board re-planning/scheduling and execution of real-time contingency actions.  This sort of functionality is broadly applicable to science missions operating in all contexts but critical to missions with significant and challenging operations outside available communication windows. (E.g. Mercury Lander nighttime operations with six weeks of no ground contact. Also, Intrepid operations that have to cover an 1,800 km distance with hundreds of instrument placements in four years, where manual interventions cannot occur more than once every 6–16 km of traverse, or missions seeking to access extreme surface or subsurface terrains. \cite{elliottIntrepid,mercuryLanderVideo,SanaeMercury,EdwardsMars})

Generally, by adopting an approach where situational awareness (perception, mapping, estimation, see \cite{matthies}), hazard assessment, planning/execution, payload data analysis, science planning, and FDIR functions are moved on-board, landed missions will be more productive. As an example, results from the Self-Reliant Rover study, wherein a terrestrial rover was operated by campaign intent rather sequenced activities, showed an 80\% reduction in sols required to complete a campaign and 267\% increase in locations surveyed per week \cite{gaines2018self}. 

\textit{Deep Space Missions} Although deep-space provides a more predictable environment than that near, on, or into planetary bodies (rendezvous, proximity operations, surface mobility, below-surface access),  autonomy enables operations with reduced communication burden and allows scaling to multi-craft missions for deep-space (orbiters and flybys). Using autonomous navigation functions provides tighter turnaround loops in high dynamic environment situations where long light-times precludes ground processing to achieve required accuracy. Autonomy can enable operations in less predictable scenarios, like atmospheric aerocapture at icy giants, where turnaround time on ground-based navigation may induce additional risk, and for planetary constellation where coordinated, multi-spacecraft operations are required \cite{DuttaAerocapture,AustinAerocapture}. 

As the sensitivity of ground-based surveys improve over the next decade, the detection frequency of interstellar visitors, e.g., `Oumuamua and 2I/Borisov, and long-period comets, like C/2017 K2, is expected to increase \cite{ExobodiesBackyard}. At the same time, numerous proposals have called for flybys of distant objects such as Trojans, Jovian and Saturnian moons, trans-Neptunian objects, and Kuiper Belt objects. Both sets of missions face similar challenges: flybys of these bodies involve high relative velocities, limiting the effectiveness of ground-based navigation and science planning to the point where the mission may not be feasible. Integrating autonomous navigation, payload data processing, and planning/execution functions may be enabling for these missions \cite{DonitzOort}. In architectures with multiple spacecraft, e.g. a two spacecraft flyby of a NEO for bistatic radar investigation, multi-agent coordination, navigation, and planning/execution can be used to perform coordinated measurements to achieve challenging measurement objectives and maximize payload utilization based on available resources.

The Mercury, Venus, Ceres, MORIE, MOSAIC, and Persephone PMCS reports have baselined electric propulsion \cite{mercuryLanderVideo,CeresHabitability,MORIE,MOSAIC,Persephone}. Electric propulsion trajectories are non-Keplerian and use continuous thrusting to benefit from their high $I_{SP}$. However, safe mode events can result in missed thrust, risking mission success in terms of schedule and excessive propellant use to correct trajectory. For Dawn, a four-day period of missed thrust resulted in a 26-day delay to the first planned Ceres orbit; a projected seven-day missed thrust could have resulted in a $\sim$50-day delay \cite{grebow2016dawn}. Restorative fault management capability would mitigate the impact of missed thrust.

Missions in the Jovian environment, like Europa Clipper, at Mercury, like Mercury Lander, and missions during solar maxima face increased risk of radiation-induced events that can induce component resets and induce safe mode events. Suspending operations as operators intervene may be acceptable during cruise, but during critical mission phases like a Europa flyby, it could cause a mission to fail its mission requirements. Restorative fault management could enable a spacecraft to recover to an operable state sufficient to perform critical measurements necessary for meeting mission requirements.

Like EO-1, orbital mapping missions would benefit from autonomous planning/execution, being able to opportunistically detect novel signatures, e.g. methane plumes at Mars or volcanic plumes at Io, and to perform follow-up observations. Missed coverage imaging, e.g. due to an anomaly, could also be replanned on board. Multi-spacecraft monitoring, like MOSAIC or the NanoSWARM concept, could use multi-agent coordination for event tracking and map completeness \cite{garrick2015nanoswarm}. In these cases, autonomy could improve a mission's science-to-dollar ratio by increasing productivity and the value of returned data.

\subsection{Institutional Drivers of Autonomy}
There are also institutional motivations for using autonomous functions on-board spacecraft. A shift in culture that is more accepting of autonomous systems would see its benefits in relaxing constraints on deep space communication and scheduling, mission competitiveness, and achieving more with the planetary exploration budget.

\textit{Deep Space Network \& Communications Infrastructure} All missions utilizing the Deep Space Network (DSN) could be expanded and see improvements in science return using adaptive operations. The introduction of deep space SmallSat missions is affecting the roadmap for NASA’s Deep Space Network (DSN) and is leading to a fundamental change in the way future deep space missions will interact with ground systems \cite{DeutschDSN}. With the projected increase in the number of DSN users, e.g. SmallSats funded through SIMPLEx, the DSN will need to more efficiently schedule tracking passes. Increasing the degree of spacecraft autonomy will allow improvements in the efficiency of DSN use.

\textit{Mission Competitiveness} The interrelated factors of cost, perceived risk, and science return impact selection of competed missions. Integrating different autonomous functions can afford PIs more flexibility in meeting science requirements, relaxing system requirements on mission systems, and potentially reducing cost and science risk. For instance, using autonomy like autonomous targeting can enable performing novel and opportunistic measurements, raising a mission's threshold science return. Use of autonomy like autonomous retry with on-board planning/execution, can reduce mission risk by allowing a spacecraft to dynamically adopt to on-board anomalies and reattempt measurements.

A notable example of a mission that could become more competitive should systems autonomy be used is a New Frontier Venus lander - a concept that has seen multiple failed proposals over two decades, including the step-two Venus In situ Composition Investigations (VICI) and the Venus Surface and Atmosphere Geochemical Explorer (SAGE) proposals \cite{glaze2017vici, jones2003surface}. By integrating the autonomous functions in Table 1, a Venus lander could more effectively utilize its time before end-of-mission. Contextual images and remote Raman/LIBS measurements can direct sampling, resulting in lower risk on science return.  Based on current development of the NASA ARC Volatiles Investigation Polar Exploration Rover (VIPER), this could enable higher performance and better handling mission-ending faults.

\textit{Cost}  Autonomy could reduce costs to NASA by moving some of operations team functions on board. In the last 10 years, SMD has spent \$2.4B FY20 on mission operations \cite{PlanetarySociety}. If these missions had been operated 20\% more efficiently, \$480M could have been available to NASA. As predicted in the 1980 Sagan Report, moving to ``Autonomous Mission Control'' by the year 2000 would result in mission operations costs 1\% of those in 1975. While “Autonomous Mission Control” has yet to be realized, the experience from past missions and current research supports that prediction. On EO-1, significant mission time was saved by having the spacecraft discard imagery obscured by cloud cover. 

\section{Why Autonomy Isn’t Standardized}
Despite decades of incremental progress in achieving remarkable successes with the autonomous functions used on EO-1, different Mars rovers, and missions like Deep Impact and Stardust, the broad use of autonomous functions to achieve mission objectives is still not standard practice. While the reasons are interconnected, they can be generalized as three barriers to infusion.

\textbf{Barrier 1: Unlike most other NASA technology investments, autonomy is system-level technology.} Most technologies, e.g. detectors and propulsion systems, fulfill specific needs. Their operational conditions can be readily modeled for laboratory testing and interfaces and behavior well-defined for mission development and operations. For these technologies, the NASA definitions of Technology Readiness Level (TRL) are a relevant and useful tool for appraising flight-readiness. Moreover, incremental investments that raise TRL are appropriate. Autonomy, especially where multiple functions are integrated, has implications for the system's architecture, design, development and operations processes, and personnel throughout. It requires coordinated efforts throughout development phases and across domains and cannot be comprehensively adopted through incremental investments. This means standard paths to mature technologies do not apply to autonomous systems. Experience has shown that technology demonstration missions, like the Remote Agent Experiment on Deep Space-1, do not transition to routine science mission use \cite{bernard1999spacecraft}.

\textbf{Barrier 2: Institutional environment limits autonomy to incremental maturation and thus restricts NASA’s ability to deploy autonomous systems.} Even though autonomy has the potential to reduce mission risk, using it is still a \textit{perceived} risk. NASA has successfully executed missions without significant autonomy for decades and is capable of developing and evaluating missions that are enhanced or enabled by autonomy. As exemplified by AutoNav and AEGIS and described in the Sagan Report and ESA OPS-SAT paper, a cultural barrier limits opportunities to fly enhancing technologies. Risk aversion results in incremental maturation of specific autonomous functions, which struggle again to find their place on science missions despite past success and potential advantages. Even when risks are taken on competed missions and they would stand to benefit in science-to-dollar ratio, e.g. Dragonfly, opportunistic science is not considered an intrinsic component of baseline missions. As also noted in the Sagan Report and still true today, NASA struggles to pioneer software advances. The nature of the competitive mission process drives engineers at NASA and in industry toward heritage solutions in mission proposals to avoid perceived risk. Integrated over time, this steers missions away from innovative advances that could be enabling for more demanding missions. There is little interdisciplinary cooperation to offer guidance on performing software- and autonomy-related trade studies, which could have significant implications for mission architecture. This lack of exposure and strained mapping of TRL to multi-mission autonomous functions further reinforces a culture wary of adopting more advanced autonomy. 

\textbf{Barrier 3: Lack of Inter-Directorate Coordination.} The Human Exploration Directorate (HEOMD), the Space Technology Mission Directorate (STMD), and divisions of Science Mission Directorate (SMD) have struggled to coordinate and lack direct incentive to develop NASA-wide plans to implement autonomous systems. This results in fragmented and incremental progress that will not lead the agency to sustainable advances in multi-mission autonomy. Alternatively, coordinated investments could allow directorates to leverage developments by others, promoting agency-wide sharing of standards and practices, increasing the likelihood autonomous functions are reflown, and minimizing duplicated investments. Meanwhile, private industry benefits from investments in autonomy and can push forward with increasingly autonomous solutions, e.g. the Falcon fly-back boosters used by SpaceX.

\section{Recommendations}
Given how autonomy would enable and enhance strategic missions of interest to the planetary science community, we offer the following high-level recommendations for enabling the routine deployment, and continued evolution, of autonomy for future planetary science missions. These high-level recommendations include several potential implementations described as examples.  

\begin{enumerate}
    \itemsep0em
    \item \textbf{Create programmatic pathways that prepare integrated autonomy systems for future missions and build institutional trust, e.g.:}
    \begin{itemize}[leftmargin=*]
        \itemsep0em
        \item Commit to advancing autonomy by setting a series of capability deadlines to include increasing amounts of autonomy on all planetary science missions. This would ensure that NASA is ready with the needed processes and capabilities when the time comes to fly the more ambitious missions that are enabled by autonomy
        \item Incentivize adoption of autonomous systems for PSD Announcements of Opportunity (AOs), e.g. through a cost incentive.
        \item Coordinate STMD investments with competed missions, e.g. SIMPLEx, so missions push boundaries of science exploration and technology demonstration
        \item Expand programs like the New Frontiers Homesteader Program and ROSES to include autonomous functions, system integration, operator tools, and verification methods for autonomous systems. \textit{(See Rec \#2)}
        \item Instrument AOs for Flagships should offer an opportunity for collaborative proposals with complementary instruments using autonomy for payload data processing and science planning
        \item Use all extended missions for demonstrating autonomous functions
    \end{itemize}
    \item \textbf{Invest in infrastructure for developing and supporting autonomous systems in space, e.g.:}
    \begin{itemize}[leftmargin=*]
        \item With inter-directorate coordination, invest in an in-space autonomy testbed, potentially utilizing SmallSats, so NASA centers and industry can test and flight validate flight and ground software and train on new processes \textit{(See Rec \#3)}
        \item Make DSN and Advanced Multi-Mission Operations System (AMMOS) investments that support the anticipated growth of customer missions and use of autonomous systems
        \item Expand Homesteader and ROSES opportunities to cover interdisciplinary autonomous research and development \textit{(See Rec \#1)}
        \item Set specific objectives and a time frame for transitioning AMMOS away from {\em a priori} planning (e.g., change from time-based sequencing to goal-based commanding). Set a date where all new missions would be expected to use this new paradigm.
    \end{itemize}
    \item \textbf{Invest in practices that promote the multi-mission use of autonomous systems. Practices includes design, development, test, verification, and operations processes and standards, e.g.:}
    \begin{itemize}[leftmargin=*]
        \item Develop common architectural patterns, principles and standards to enable confident integration of autonomy technologies; invest in updates to development and operations processes to enable the trustworthy deployment of increasingly autonomous missions
        \item Update mission selection and review processes to consider assessment of agency risk posture, e.g., the risk of not including new technology or methods in NASA planetary science missions.
        \item With inter-directorate coordination, invest in laboratory and virtual autonomy testbeds so NASA centers and industry can test and validate autonomy software and train on new processes. \textit{(See Rec \#2)}
        \item Spur the adoption of integrated autonomous functions in industry to support NASA’s competed missions. \textit{(See Rec \#1)}
        \item Adopt new maturity evaluations for software and model trust to augment TRL in assessing autonomous functions, their integration, and applications.
    \end{itemize}
    \item \textbf{Set goals to reduce operations costs and determine the degree of autonomous operations required to achieve these goals.} NASA should commission an independent study, potentially performed by the National Academy of Sciences, assessing existing operations to evaluate how operations costs can be reduced by adopting autonomy and what funding and savings profiles would result.	  
\end{enumerate}

\section{Acknowledgements}
Thanks to Ellen Van Wyk (NASA JPL) for illustrations. 

%\section{References}
%Please see this link for the full list of references: \url{https://bit.ly/3hsLk8o}.

%\newpage
\printbibliography
\end{document}